\documentclass[ aip, jmp, preprint, showpacs,preprintnumbers,amsmath,amssymb]{revtex4-1}

\usepackage{amsmath}
\usepackage{graphicx}
\usepackage{dcolumn}
\usepackage{bm}
\usepackage{color}

\begin{document}


\title{Effects of the non-uniform initial environment and the guide field on the plasmoid instability}

\author{Lei Ni }
\email{leini@ynao.ac.cn}
\affiliation{
Yunnan Astronomical Observatory, Chinese Academy of Sciences, Kunming 650011, China
 }
 
\author{Jun Lin}
\affiliation{
Yunnan Astronomical Observatory, Chinese Academy of Sciences, Kunming 650011, China
}
\affiliation{
Harvard-Smithsonian Center for Astrophysics, 60 Garden Street, Cambridge, MA 02138, USA
}

\author{Nicholas A. Murphy }
\affiliation{
Harvard-Smithsonian Center for Astrophysics, 60 Garden Street, Cambridge, MA 02138, USA
}


\date{\today}
  

 
\begin{abstract}

Effects of non-uniform initial mass density and temperature on the plasmoid instability are studied via 2.5-dimensional resistive magnetohydrodynamic(MHD) simulations. Our results indicate that the development of the  plasmoid instability is apparently prevented when the initial plasma density at the center of the current sheet is higher than that in the upstream region. As a result, the higher the plasma density at the center and the lower the plasma $\beta$ in the upstream region,  the higher the critical Lundquist number needed for triggering secondary instabilities. When $\beta =0.2$, the critical Lundquist number is higher than $10^4$. For the same Lundquist number, the magnetic reconnection rate is lower for the lower plasma $\beta$ case. Oppositely, when the initial mass density is uniform and the Lundquist number is low, the magnetic reconnection rate turns out to be higher for the lower plasma $\beta$ case. For the high Lundquist number case ($>10^4$) with uniform initial mass density, the magnetic reconnection is not affected by the initial plasma $\beta$ and the temperature distribution. Our results indicate that the guide field has a limited impact on the plasmoid instability in resistive MHD.        
\end{abstract}

\maketitle

\section{Introduction}
Magnetic reconnection\cite{Sweet1958, Parker1957, Priest2000, Biskamp2000}  is an important and basic mechanism for the magnetic energy conversion in astrophysical and laboratory plasma systems.  It converts the magnetic energy into plasma kinetic and thermal energy, which may account for eruptive  phenomena observed both in solar\cite{Priest2000} and the other astrophysical environments\cite{Yuan2009}. Recently, plasmoids have been identified in the coronal mass ejection (CME) current sheets \cite{Song2012, Savage2010, Nishizuka2010, Milligan2010, Lin2005},  Earth's magnetotail\cite{Wang2010}, and in reconnection in laser-produced plasmas\cite{Dong2012}. In resistive magnetohydrodynamic(MHD), many numerical simulations\cite{Bhattacharjee2009, Huang2010, Shen2011, Barta2011} demonstrate that the reconnection process is dominated by the secondary plasmoid instability as the Lundquist number exceeds a critical value. The reconnection rate reaches a high value $\sim 0.01$ as large numbers of plasmoid form in the current sheet\cite{Bhattacharjee2009, Huang2010}. 

The critical Lundquist number for the occurrence of secondary instabilities is estimated be around $10^4$ according to Biskamp's analysis\cite{Biskamp2000}.  However, one of our previous reduced MHD simulations\cite{Ni2010} with incompressible plasma demonstrated that this critical value is around $2 \times 10^3$.  Recently\cite{Ni2012}, by solving the 2D compressible MHD equations with different plasmas $\beta$ and Lundquist number, we found that the critical Lundquist number depends on the initial upstream plasma $\beta$, it is around $2 \times 10^3 \sim 3 \times 10^3$ for $\beta = 50$ and $8 \times 10^3 \sim 10^4$ for $\beta=0.2$. The average reconnection rate, normalized to the asymptotic value of upstream $BV_A$ is lower in a low $\beta$ system than that in a high $\beta$ system. Since no guide field was included in those simulations and the initial temperature was uniform in all those models, the low $\beta$ at the inflow upstream results in a nonuniform density distribution in the direction that is vertical to the current sheet. Therefore, the $\beta$-dependence mentioned above may be largely attributed to the nonuniform initial density distribution.  

Based on our previous work\cite{Ni2012}, the effects of the initial plasma $\beta$ on the magnetic reconnection process with plasmoid instabilities are comprehensively studied in this paper. Three models with different guide fields, initial plasmas mass density and temperature distributions are introduced. Simulations with different Lundquist number and different initial plasma $\beta$ in the asymptotic inflow region have been performed in the three models. The characteristics of the current density, the magnetic flux,  the reconnection rate and the magnetic energy spectrum along the current sheet during the evolutionary processes of  the plasmoid instability are studied. Numerical simulations were carried out  with the MHD code NIRVANA version 3.4\cite{Ziegler2008} in the 2.5-dimensional Cartesian space.     

In next section, the MHD equations governing the evolution in the system, together with the associated initial conditions are described in detail. In section III, we present the main numerical results in the three models. Discussions and a summary  are given in section IV.        
       
 \section{Governing Equations and Initial Conditions}

As we have described in our previous work\cite{Ni2012},  the dimensionless MHD equations we used in the code are as below:
\begin{equation}
 \partial_t \rho = -\nabla \cdot (\rho \textbf{v}),
\end{equation}

\begin{equation}
 \partial_t e = -\nabla \cdot [(e+p+\frac {1}{2 }\vert \textbf{B} \vert^2)\textbf{v}- (\textbf{v} \cdot \textbf{B})\textbf{B}] + \eta \nabla \cdot [\textbf{B} \times
 (\nabla \times \textbf{B})],
\end{equation}

\begin{equation}
\partial_t (\rho \textbf{v}) = -\nabla \cdot [\rho \textbf{v} \textbf{v} +(p+\frac {1}{2} \vert \textbf{B} \vert^2)I - \textbf{B} \textbf{B} ],
\end{equation}

\begin{equation}
\partial_t \textbf{B} = \nabla \times (\textbf{v} \times \textbf{B})+ \eta\nabla^2\textbf {B},
 \end{equation}
 
 \begin{equation}
 e = p/(\Gamma_0-1)+\rho \textbf{v}^2/2+\textbf{B}^2/2.
 \end{equation}
 
The variables above are only functions of space in $(x, y)$ direction and time $t$. The simulation domain is from $0$ to $1$ $(l_x=1)$ in $x$-direction and from $0$ to $4$ ($l_y=4$) in $y$-direction. The Lundquist number is defined as $S=l_y v_A/\eta$, where $v_A$ is the initial asymptotic Alfv\'en speed in the upstream boundary, which is equal to unity in our calculations. We use three models to describe the initial conditions for our simulations.  Each model is incorporated into several different simulations with different Lundquist number and initial plasma $\beta$ at the $x$ boundary.

 In model I,  we start with a Harris sheet in the $(x, y)$ plane and a uniform guide field in the $z$-direction:
 \begin{equation}
  B_{x0} = 0, \quad B_{y0} = b_{y0} \tanh(\frac{x-0.5}{\lambda}), \quad B_{z0} = b_{z0},
\end{equation} 
where $\lambda$ is the width of the Harris sheet and is  set equal to $0.05$, which is small enough to allow tearing instabilities to develop\cite{Ni2012}. We choose $b_{y0} = 0.8$ and $b_{z0} =0.6$ in this model. The initial configuration is in both thermal and mechanical equilibrium, so the initial velocity is zero. From equation(3), the plasma pressure satisfies the initial equilibrium condition:

\begin{equation}
  \nabla \cdot (p_0 \textbf{I}) = -\nabla \cdot [\frac{1}{2}\vert \textbf{B}_0 \vert^2   \textbf{I}- \textbf{B}_0\textbf{B}_0].
\end{equation}
 
 Since $\textbf{B}_0=B_{y0} \hat{\textbf{y}} + B_{z0} \hat{\textbf{z}}$, where $\hat{\textbf{y}}$ is the unit vector in $y$-direction and $\hat{\textbf{z}}$ is the unit vector in $z$-direction,  the initial equilibrium gas pressure is calculated as:
\begin{equation}
  p_0 = -\frac{1}{2}(B_{y0}^2 + B_{z0}^2)+C_0, 
\end{equation}  
where $C_0$ is a constant. From equation (6), we know that $B_{y0}^2 + B_{z0}^2 = 1$ at the $x$ boundary. Since  the plasma gas pressure is related to the magnetic pressure by $\beta = 2p/B^2 $ ,  we get $C_0= (\beta_0 +1)/2$, where $\beta_0$ is the initial plasma $\beta$ at the $x$ boundary.  Thus:
\begin{equation}
   p_0 = \frac{1+\beta_0-B_{y0}^2-B_{z0}^2}{2}.
\end{equation}
The initial equilibrium state of the total energy is:
\begin{equation}
    e_0 = p_0/(\gamma -1)+(B_{y0}^2+B_{z0}^2)/2.
\end{equation}
From the ideal gas law $T=p/\rho$,  and the assumption of a uniform temperature, the initial equilibrium mass density and temperature are:
\begin{equation}
  \rho_0 = p_0/T_0=\frac{1+\beta_0-B_{y0}^2-B_{z0}^2}{\beta_0}, \quad T_0=\frac{\beta_0}{2},
\end{equation}
respectively. As such, in model I, the initial distributions of gas pressure and mass density depend on the plasma $\beta_0$ at the inflow boundary. The lower the $\beta_0$, the larger the gradient of the mass density from the center to the inflow boundary.

In model II, the initial distributions of the magnetic fields and gas pressure are the same as in model I, but we use a non-uniform initial distribution of the temperature, which varies with $\beta_0$. The initial mass density is assumed uniform. Therefore, the distributions of the initial equilibrium temperature and mass density in model II are:
\begin{equation}
   T_0 = p_0/\rho_0=\frac{1+\beta_0-B_{y0}^2-B_{z0}^2}{2}, \quad \rho_0 = 1.0 .
 \end{equation}
 The lower the $\beta_0$, the larger the gradient of the temperature from the center to the inflow boundary. 
 
In model III, we use nonuniform guide fields in $z$-direction. The distributions of three components of  the initial equilibrium guide field are given as below:
\begin{equation}
   B_{x0} = 0, \quad B_{y0} = b_{0} \tanh(\frac{x-0.5}{\lambda}), \quad B_{z0} = b_{0}/\cosh(\frac{x-0.5}{\lambda}),
\end{equation}
where $b_0 = 1.0$. From these expressions, one can notice that $B_{x0}^2+B_{y0}^2+B_{z0}^2=1.0$.  The width of the Harris sheet $\lambda$ is still set to equal $0.05$ in this model. As we have described in model I, from equation(7), we can get the initial pressure:
\begin{equation}
   p_0 = \frac{1+\beta_0-B_{y0}^2-B_{z0}^2}{2}=\beta_0/2.
\end{equation}
Hence, the initial equilibrium gas pressure is uniform, and the initial equilibrium state of the total energy is:
\begin{equation}
   e_0 = p_0/(\gamma -1)+1/2.
\end{equation}
In this model, we assume that both the initial equilibrium density and temperature are uniform: \begin{equation}
  \rho_0 = 1.0, \quad T_0=\frac{\beta_0}{2}.
 \end{equation}
Therefore, the lower the $\beta_0$, the lower the temperature and gas pressure in the whole simulation domain.  

In Fig.1, we present the initial equilibrium conditions along the $x$-direction in the three models for the case of $\beta_0=0.2$ and $\beta_0=50$. In all three models, the following perturbation is added to the magnetic field:  
  
\begin{equation}
   b_{x1} = -\epsilon \cdot 0.5 \sin(\pi x/l_x)\cos(2\pi y/l_y),
 \end{equation}
                        
\begin{equation}
   b_{y1} = \epsilon \cdot \cos(\pi x/l_x)\sin(2\pi y/l_y).
\end{equation}
A constant value $\epsilon = 0.05$ is  used in our simulations, which is the same as that used in  our previous paper$^{[15]}$.  We use periodic boundary condition in $y$-direction and Neumann boundary condition in $x$-direction. The highest Lundquist number we have tested is $2 \times 10^5$ in this work. The adaptive mesh refinement (AMR) technique is used in the code, and we start the simulation with a base level grid of $80\times 320$. The highest refinement level in our experiment is 8, which corresponds to a grid resolution $\delta_x = 1/20480$. Convergence studies have been carried out to test the cases $S=2\times10^5$ and $\beta_0 = 0.2$ in all of the three models with a lower refinement level of 7, corresponding to a grid resolution of $\delta_x=1/10240$.  The reconnection rate is similar to the higher resolution run with $\delta_x=1/20480$. Therefore, the grid resolution in our simulations is sufficiently high to suppress
 the numerical resistivity.

\section{numerical results}

As we have demonstrated clearly in our previous paper\cite{Ni2012}, the critical Lundquist number for the  onset of the plasmoid instability depends on $\beta_0$. However, this $\beta_0$ dependence could be largely attributed to the density variation. The reconnection processes with different Lundquist number and plasma $\beta_0$ have been studied numerically in all of the three different models which have been described in section II. Table 1 presents the simulations with different initial $\beta_0$ and Lundquist number $S$ that we have carried out in this paper. For example, in model I for $\beta_0 = 0.2$, we have simulated the reconnection process for  $S \in \{2 \times 10^3, 5 \times 10^3, 7 \times 10^3, 10^4, 2 \times 10^4, 2.5 \times 10^4, 3 \times 10^4, 4 \times 10^4 \} $. No secondary instabilities appear for $S \leq 2.5 \times 10^4$, and secondary plasmoids start to appear when $S \geq 3 \times 10^4$. Therefore, the critical Lundquist number is between $S=2.5 \times 10^4$ and $S=3 \times 10^4$ for $\beta_0= 0.2$ in model I. By using the same methods, the critical Lundquist number is found between $S=3 \times 10^3$ and $S=4 \times 10^3$ for $\beta_0=50$ in model I. In model II,  the critical Lundquist number is found between $S=4\times10^3$ and $S=5\times10^3$ for $\beta_0=0.2$, and this critical value becomes $3 \times 10^3 \sim 4 \times 10^3$ for $\beta_0=50$. In model III,  we find that the critical Lundquist number is around $7 \times 10^3$ to $8 \times 10^3$ for $\beta_0=0.2$, and that the  critical value for $\beta_0=50$ is between $S=3 \times 10^3$ and $S=4 \times 10^3$.  Therefore, the critical Lundquist number for the occurrence of secondary instabilities depends on the initial plasma $\beta_0$ at the inflow boundary. This critical value is usually higher in the lower $\beta_0$ case. In model I with non-uniform initial mass density,  this phenomenon is more obvious, the critical Lundquist number is around an order of magnitude higher for $\beta_0=0.2$ than that for $\beta_0=50$. In model II and III, the initial mass density are both uniform in the two models,  this critical value is no more than two times higher for $\beta_0=0.2$ than that in the $\beta_0=50$ case. 

In the following part of this section, the time dependent reconnection rate, the evolution of the current density and magnetic flux, the magnetic and the kinetic energy spectrum along the current sheet are demonstrated in different models with different $\beta_0$. The reconnection rate $\gamma$ is calculated using the same method as we have described in our previous paper\cite{Ni2012},  $\gamma = \partial (\psi_X - \psi_O)/\partial t$, where $\psi_X$ and $\psi_O$ are the magnetic flux function at the main reconnection $X$ point (where the separatrices separating the two open field line regions intersect) and the $O$ point. The method used to get the magnetic and kinetic energy spectrum $E(k) \sim k^{-\alpha}$ is also the same as we have used previously. 

Fig.3 shows the evolution of the current density and magnetic flux with $S=2 \times 10^5$ for different models. Since the current sheets are very thin for such a high Lundquist number, in order to see the details inside the current sheet clearly,  the plots are stretched in the $x$-direction and only the simulation domain from $0.4$ to $0.6$ in $x$-direction is presented. Hence, the real current sheets are much thinner than the plots presented in Fig.3.  From Fig.3(a) and Fig.3(b), one can see that the secondary magnetic islands appear earlier and the secondary current sheets are thinner for $\beta_0 = 50$ case than that for the $\beta_0 = 0.2$ in model I. The current density at the reconnection $X$-points  is also higher for $\beta_0=50$ during the later stage of the secondary instability process. In model II and III,  the whole reconnection process with secondary instabilities is very similar for the $\beta_0=50$ case and the $\beta_0=0.2$ case, we only present the results of model III in Fig.3(c) and Fig.3(d) here.  As we know, the smaller the $\beta_0$ , the higher the gradient of the mass density from the center to the inflow boundary in model I, and the higher the temperature gradient from the center to the inflow boundary in model II.  Therefore, the effects of the initial $\beta_0$ on the secondary plasmoid instability with high Lundquist number  is essentially decided by the distribution of the initial mass density. These results indicate that the non-uniform distribution of the initial mass density in $X$-direction can strongly affect the secondary plasmoid instability.  The influence of the initial temperature on this process, however, is not apparent.  
 
For $\beta_0 = 0.2$ and $S=2 \times 10^5$, the time dependent magnetic energy spectral index $\alpha$ along the current sheet in the three models is presented in Fig.4. The value of $\alpha$ is calculated using the same way as we have described in our previous paper: the magnetic  field components $B_x$, $B_y$ and $B_z$  along the reconnection layer in the center ($x=0.5$) are selected. They are then transformed to Fourier space as $\tilde{B_x}(k)$, $\tilde{B_y}(k)$, $\tilde{B_z}(k)$, and the magnetic energy spectrum $E_{B}(k)$ is obtained as $E_{B}(k)\equiv  (\tilde{B_x}^2(k)+\tilde{B_y}^2(k)+\tilde{B_z}^2)/2$.  Finally, we fit the power spectrum $E_{B}(k)$ to a power law  $E_{B}(k) \sim k^{-\alpha}$  to obtain the magnetic energy spectral index $\alpha$. We only choose the region before $E_{B}(k)$ drops to a value that is five magnitudes smaller than the maximum value. For example, if the maximum $E_{B}(k)$  is $10^{-3}$,  we just fit a line to get the spectral index $\alpha$ within the region $10^{-3} < E_{B}(k) < 10^{-8}$.  Because the spectrum does not necessarily follow a power law,  the value of $\alpha$ we get here is an average value at each time step. Fig.4 shows that the spectral index $\alpha$ decreases with time,  eventually they settle down to approximately 2 in all the three models. As we have found in our previous paper\cite{Ni2012}, the value of $\alpha$ measures how smooth the current sheet is. When $\alpha$ starts to decrease, it means that the finer structures begin to appear inside the reconnection layer. Our results show that $\alpha$ decreases faster in model II and III than in model I,  which is consistent with the phenomena we observed in Fig.3(a) and Fig.3(c) that the secondary current sheets start to appear earlier in Model III than in Model I for the same $\beta_0$.  Fig.5 presents the magnetic and kinetic energy spectrum before and after secondary islands appear in model I for $S=2\times 10^5$, $\beta_0=0.2$.  At $t=2.2 t_A$, the spectral index for both magnetic and kinetic energy is high, which means that these two kinds of energy can only be cascaded to a large scale at this time point. At $t=11t_A$, the secondary islands already appear,  the value of $\alpha$ is only around $2$ for the magnetic energy spectrum and $3$ for the kinetic energy spectrum. The spectral index is set by the formation of islands on small scales, the growth of islands through 

 Different from the model in our previous paper, the guide field is included in all the three models in this work, and the effect of the initial non-uniform and uniform temperature distributions on the plasmoid instability have been studied systematically here.  Though the guide field in model II and model III is different, the numerical experiment results for these two models are very similar. The above results for Model I are also very similar to some results in the pure two dimensional  model\cite{Ni2012}. Therefore, in the MHD scale, we can conclude that the effect of the guide field is not significant in the plasmoid instability process we have studied here. However, these guide field could be very important for particle acceleration\cite{Hamilton2003, Litvinenko1996, Li2012}. Guide fields have been found to be very important during kinetic simulations\cite{Daughton2006, Drake2006, Huang2011}, which have shown that the strength of the guide field controls whether or not secondary magnetic islands can appear during magnetic reconnection\cite{Drake2006}.
 
\begin{table}
 \caption{Summary of the key parameters (plasma $\beta_0$ and the Lundquist numbers) used in the conducted numerical simulations in three different models.}
 \label{models}  
 \begin{tabular}{|l|c|l|c|l|} 
 \hline 
                &  $\beta_0= 0.2$  & $S=2000, S=5000, S=7000, S=10^4, S=2 \times 10^4$ \\
                &                               & $S=2.5 \times 10^4, S=3 \times 10^4, S =4 \times 10^4,  S= 2 \times  10^5$ \\ \cline{2-3} 
  ModelI  &  $\beta_0= 1 $    & $S= 2000, S=3000, S=4000, S=5000, S=2 \times 10^5 $\\  \cline{2-3}
                &  $\beta_0= 5 $     & $S= 2000, S=3000, S=4000, S=5000, S=2 \times 10^5 $ \\ \cline{2-3}
                &  $\beta_0= 50$    & $S=2000, S=3000, S=4000, S=5000, S=2 \times 10^5  $\\  \cline{2-3}
\hline \hline 
                &  $\beta_0= 0.2$  & $S=2000, S=3000, S=4000, S=5000, S=6000$ \\
                &                               & $S=7000, S= 8000, S = 1 \times 10^4,  S= 2 \times 10^5$ \\ \cline{2-3} 
  ModelII &  $\beta_0= 1$     & $S= 2000, S=3000 $ \\  \cline{2-3}
                &  $\beta_0= 5 $     & $S= 2000,S=3000, S=4000, S=5000, S=2 \times 10^5 $         \\ \cline{2-3} 
                &  $\beta_0= 50$    & $S=2000, S=3000, S=4000, S=5000 $  \\  \cline{2-3}
\hline \hline 
                &  $\beta_0= 0.2$  & $S=3000, S=5000, S=7000, S=8000, S=9000$ \\
                &                                & $S=1 \times 10^4, S=1 \times 10^4, S = 2 \times 10^4,  S= 2 \times 10^5$ \\ \cline{2-3} 
  ModelIII&  $\beta_0= 1$     & $S= 2000, S=3000 $\\  \cline{2-3}
                &  $\beta_0= 5$      & $S= 2000, S=3000, S= 4000, S=5000, S=2 \times 10^5 $ \\ \cline{2-3}
                &  $\beta_0= 50$    & $S= 2000, S=3000, S=4000, S=5000 $  \\  \cline{2-3}
 \hline 
\end{tabular} 
\end{table}

\begin{figure*}
\includegraphics[width=0.4\textwidth, clip=] {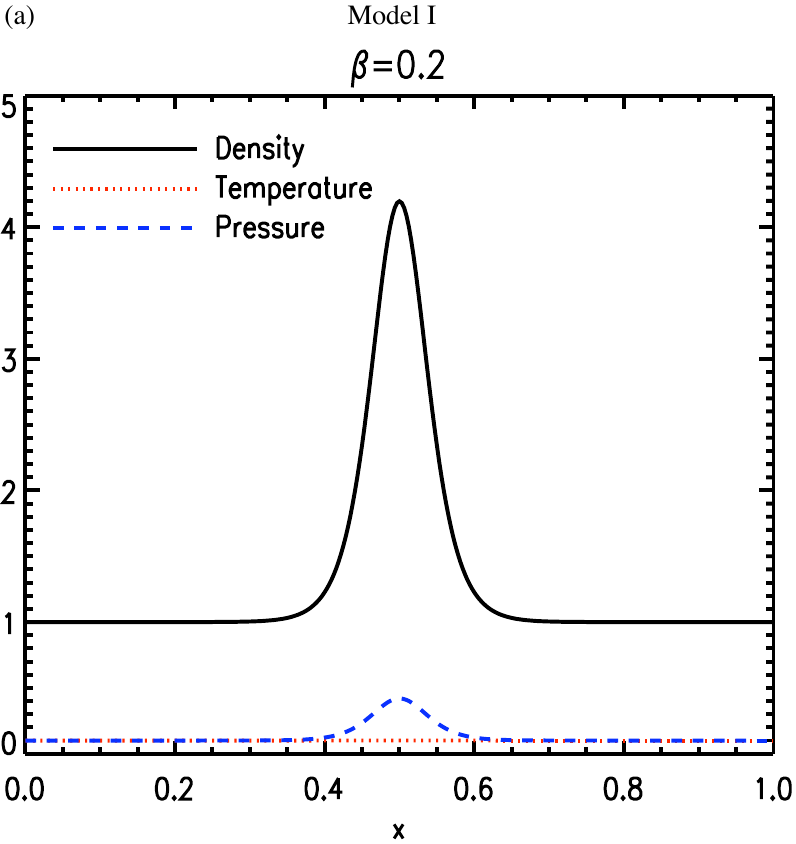}
\includegraphics[width=0.4\textwidth, clip=]  {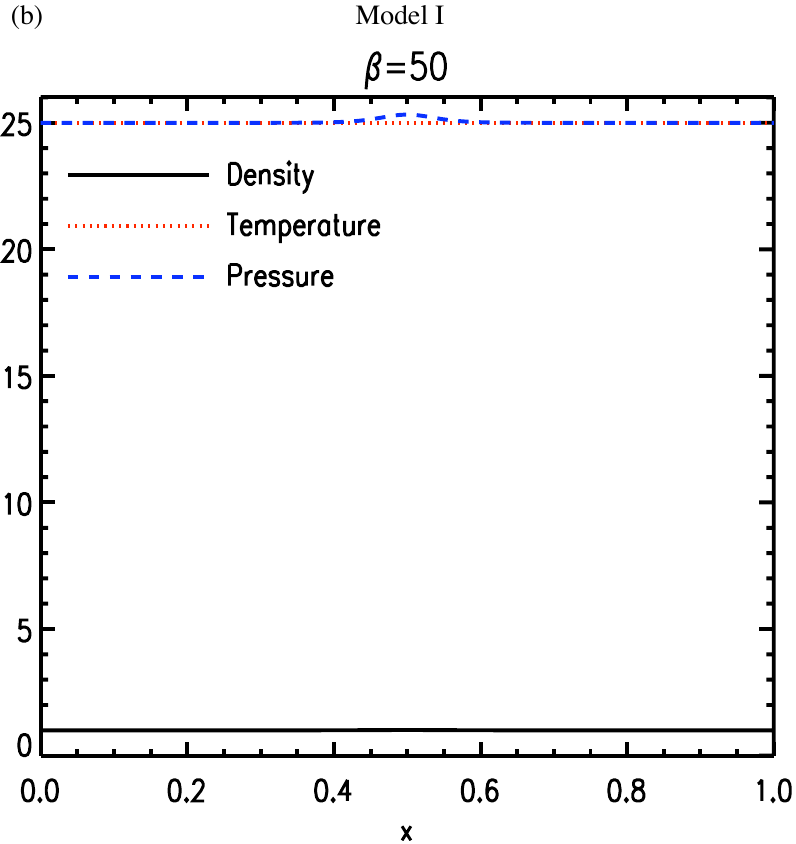}
\includegraphics[width=0.4\textwidth, clip=]  {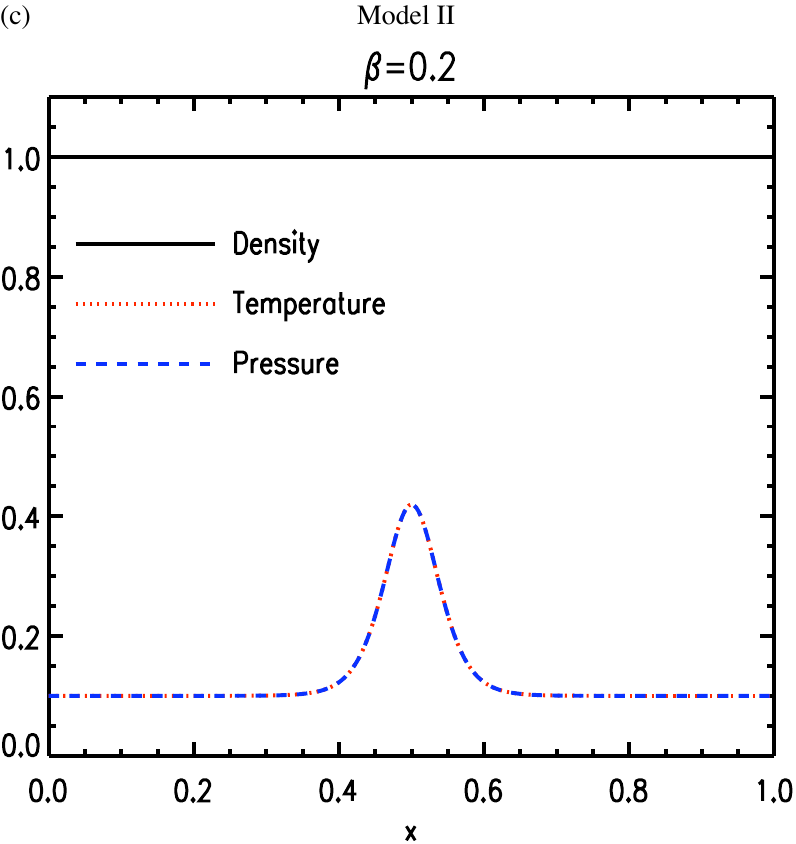}
\includegraphics[width=0.4\textwidth, clip=]  {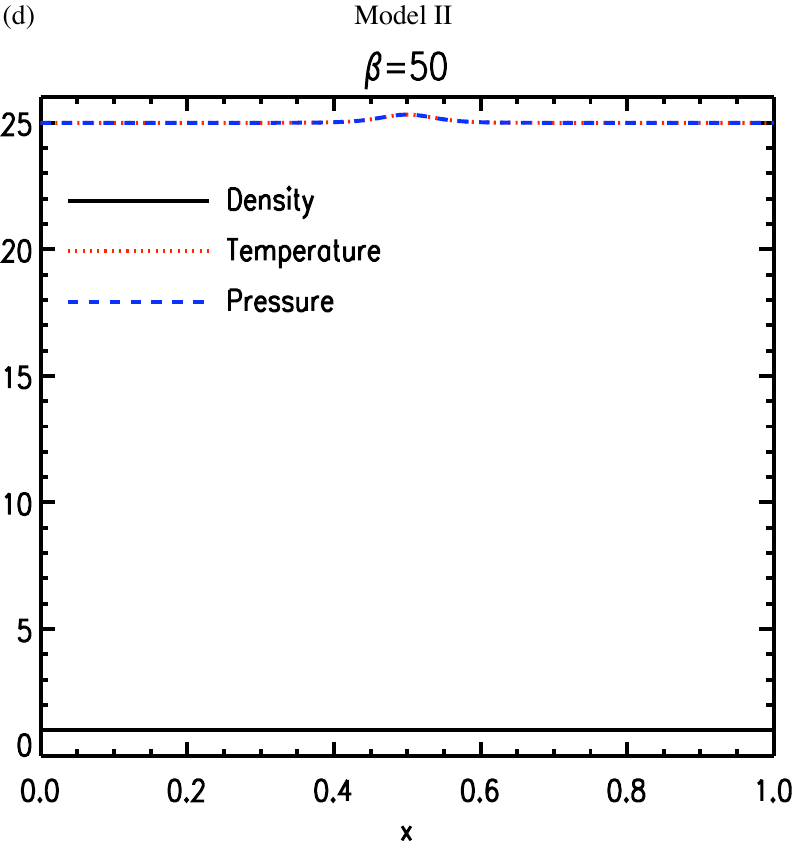}
\includegraphics[width=0.4\textwidth, clip=]  {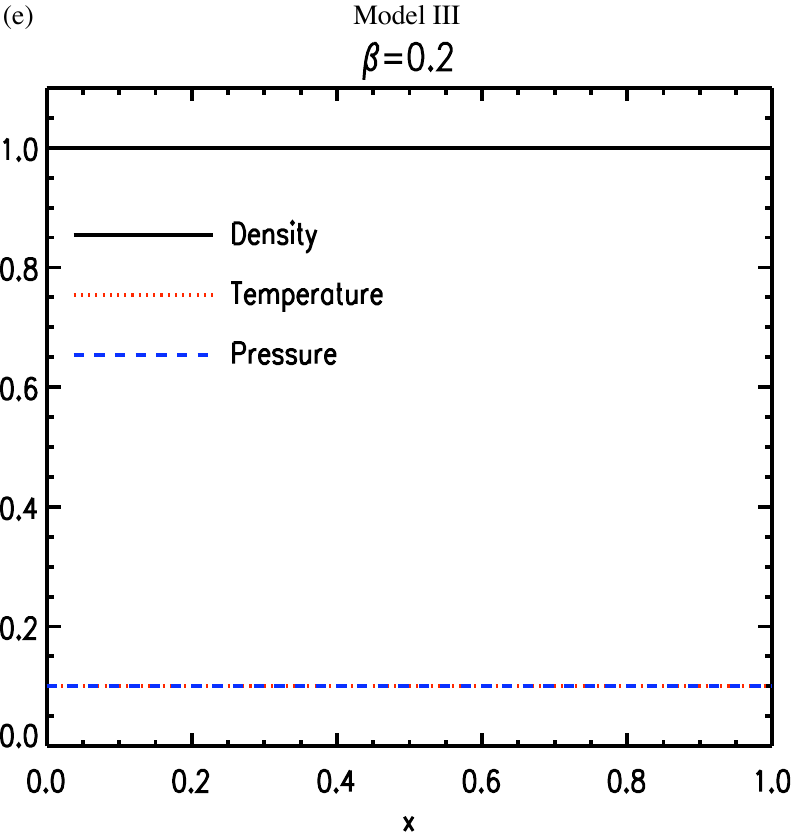}
\includegraphics[width=0.4\textwidth, clip=]  {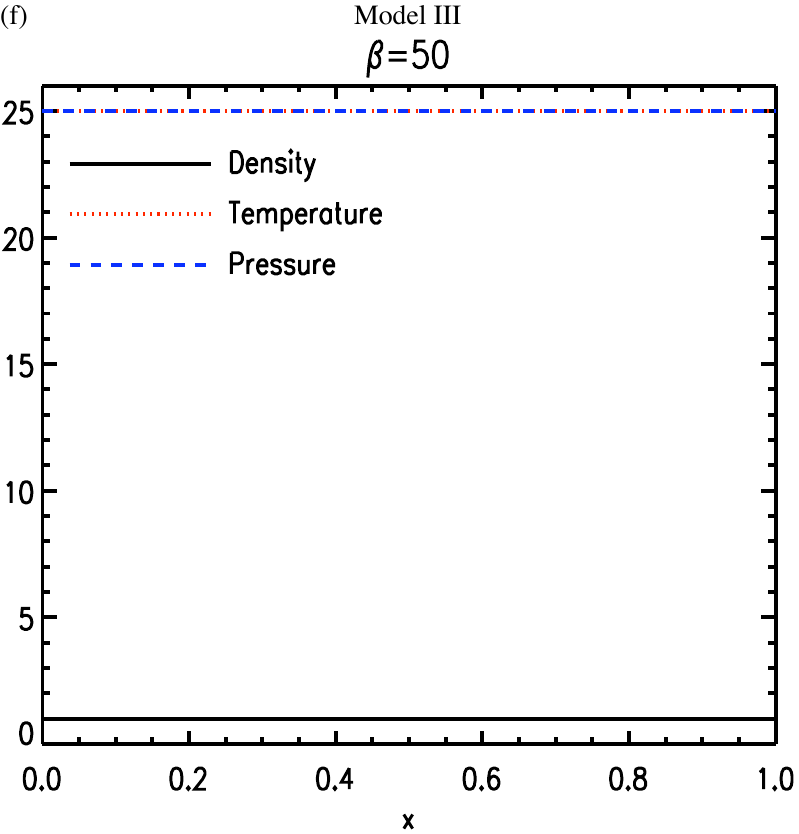}
\caption{\label{fig:1} The initial distribution of density, temperature and pressure in $x$-direction for different models with $\beta_0=0.2$ and $\beta_0$=50}
\end{figure*}

\begin{figure*}
\includegraphics[width=0.4\textwidth, clip=] {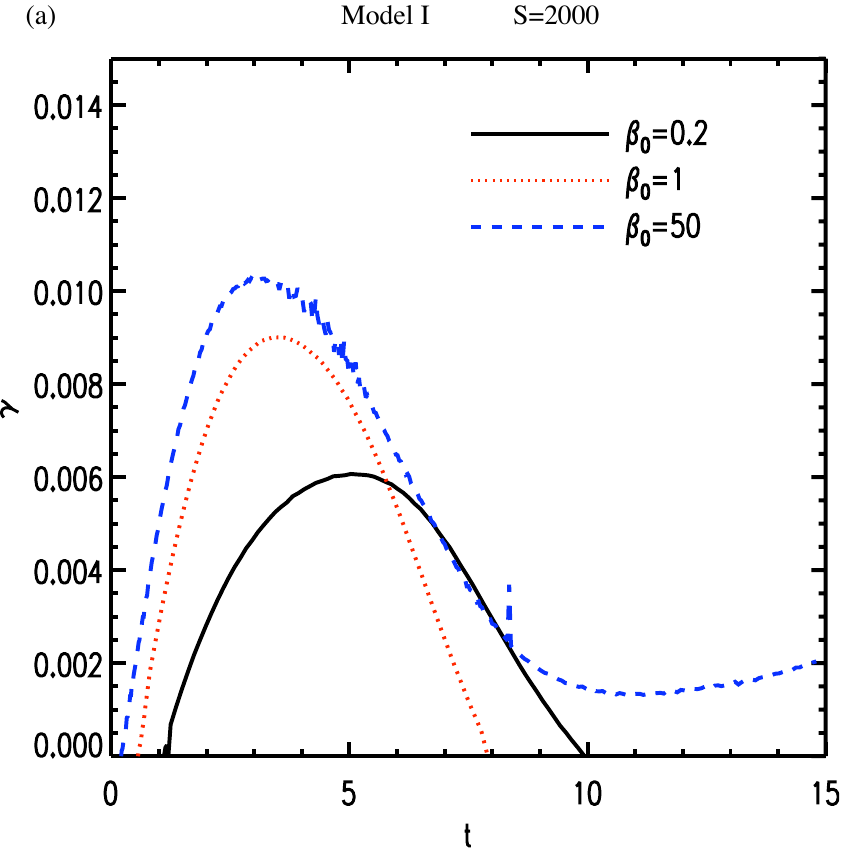}
\includegraphics[width=0.4\textwidth, clip=]  {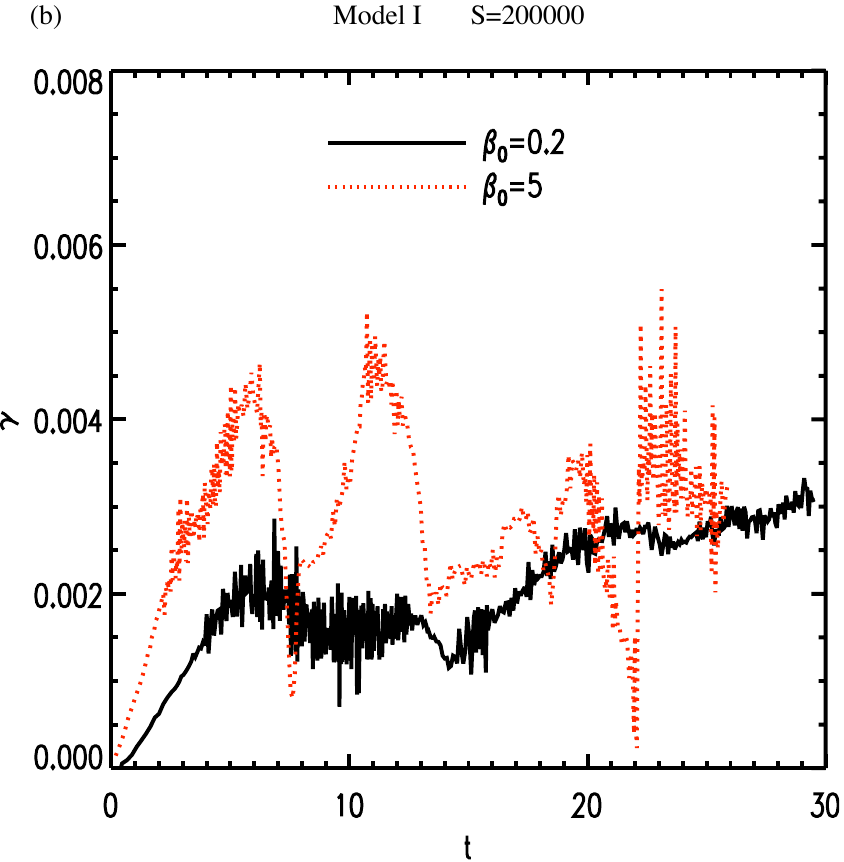}
\includegraphics[width=0.4\textwidth, clip=]  {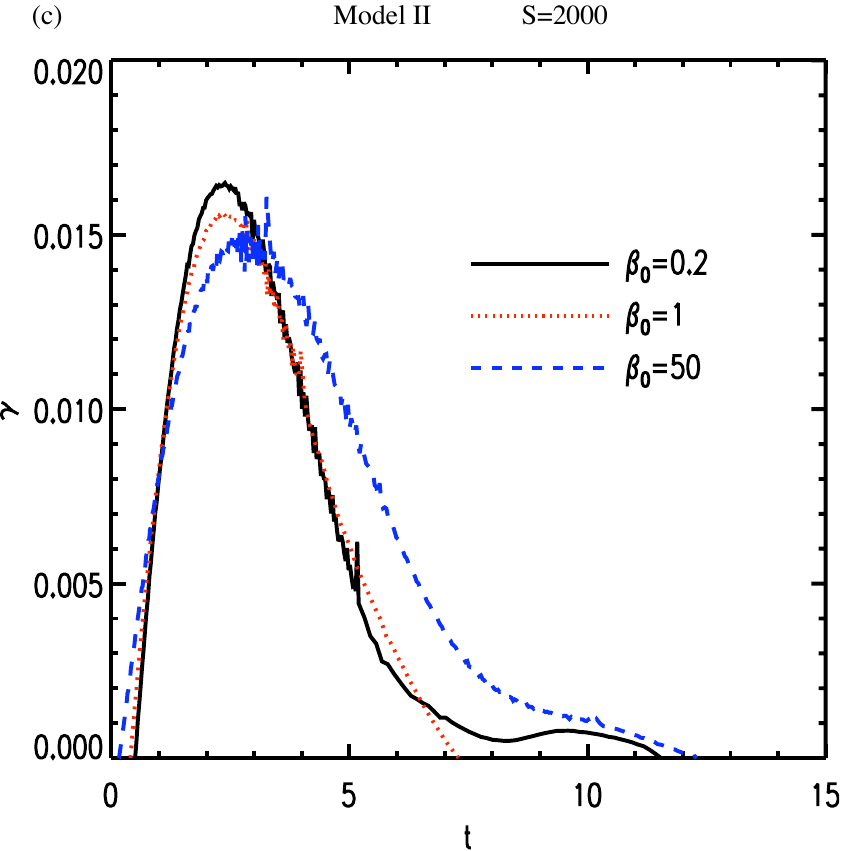}
\includegraphics[width=0.4\textwidth, clip=]  {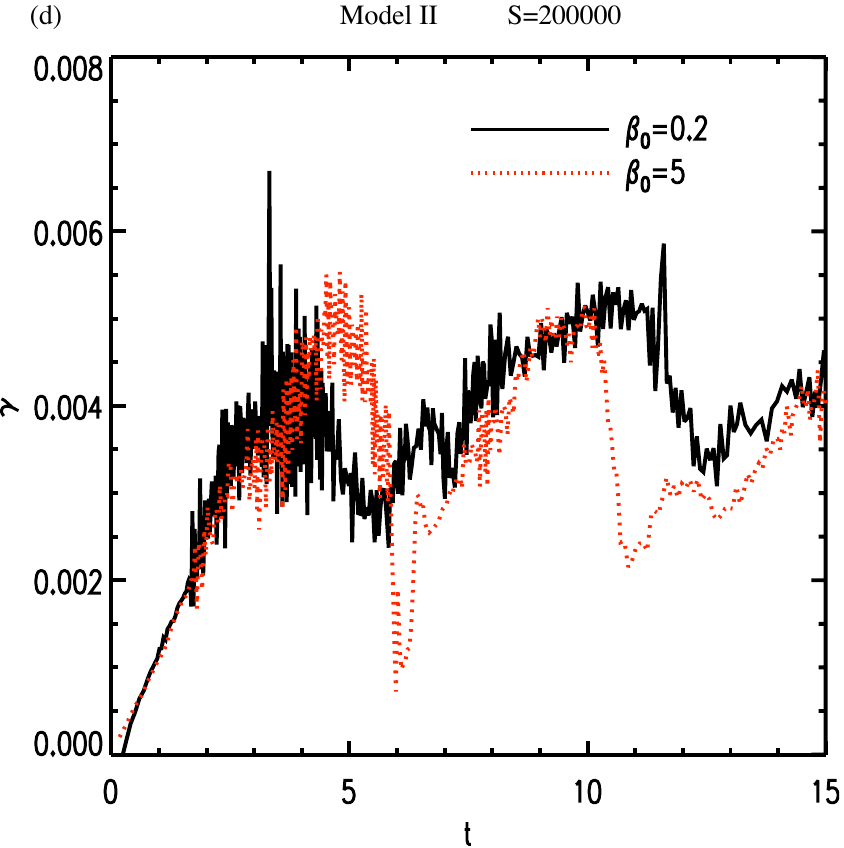}
\includegraphics[width=0.4\textwidth, clip=]  {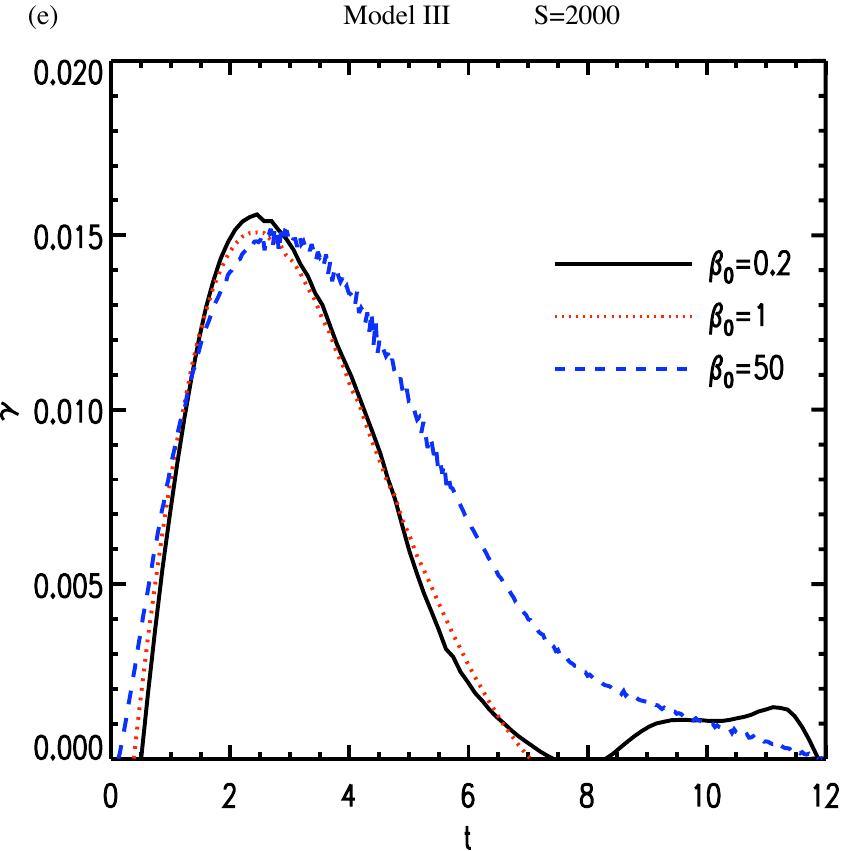}
\includegraphics[width=0.4\textwidth, clip=]  {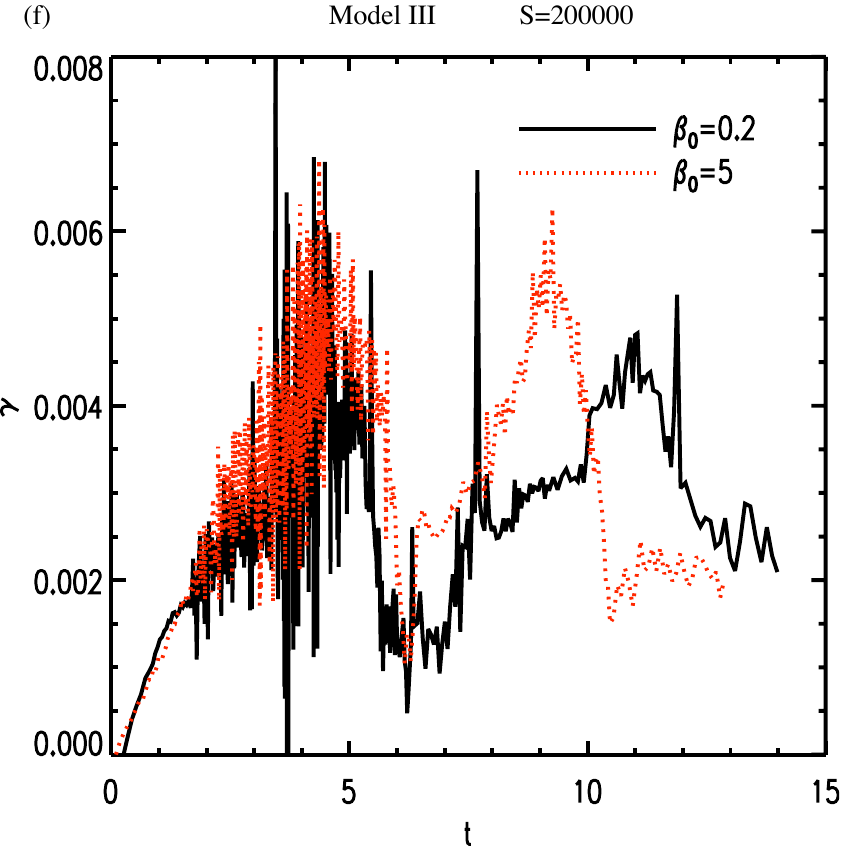}
\caption{\label{fig:2} The time dependent magnetic reconnection rate in the three models for different Lundquist number and plasma $\beta_0$.  }
\end{figure*}

\begin{figure*}
\includegraphics[width=0.55\textwidth, clip=] {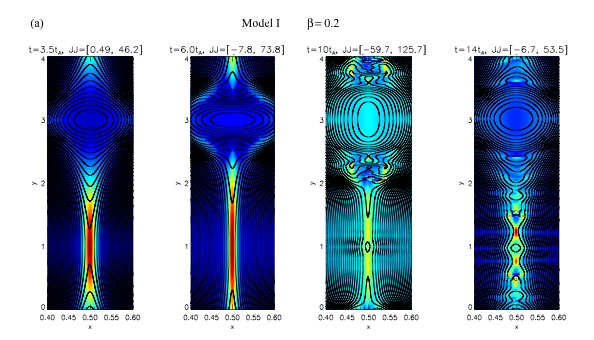}
\includegraphics[width=0.55\textwidth, clip=] {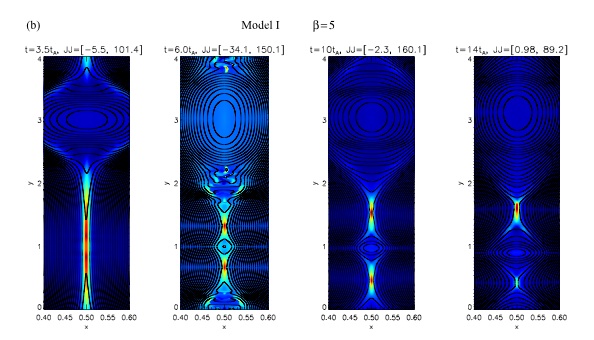}
\includegraphics[width=0.55\textwidth, clip=] {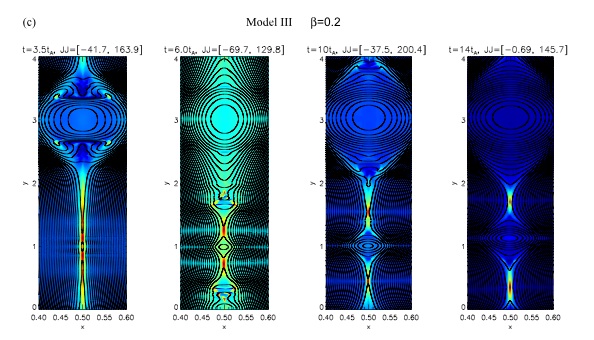}
\includegraphics[width=0.55\textwidth, clip=] {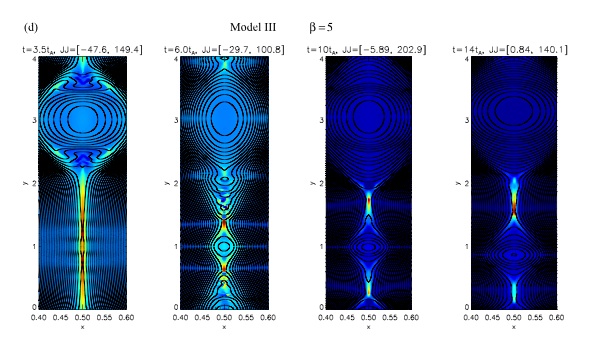}
\caption{\label{fig:3} The evolution of the magnetic flux (black line) and current density (the color plot) for $S=2 \times 10^5$ in model I and model III with $\beta_0=0.2$ and $\beta_0=5$.  }
\end{figure*}

\begin{figure}
\includegraphics[width=0.6\textwidth, clip=] {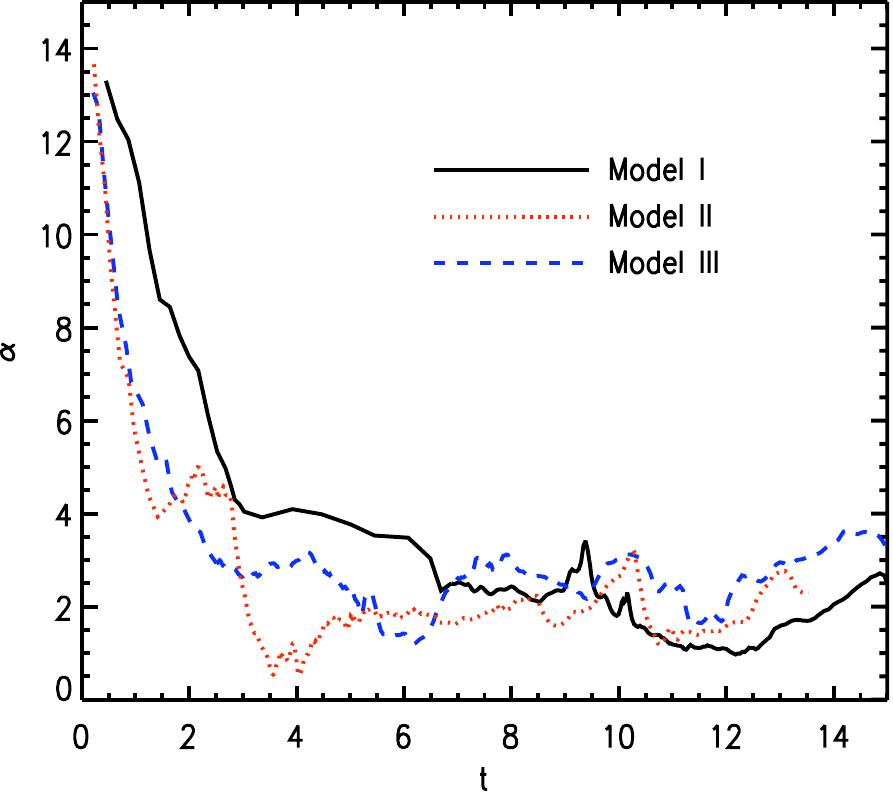}
\caption{\label{fig:4} The magnetic energy spectral index evolves with time in different models with $\beta_0=0.2$ and $S=2 \times 10^5$. }
\end{figure}

\begin{figure*}
\includegraphics[width=0.45\textwidth, clip=] {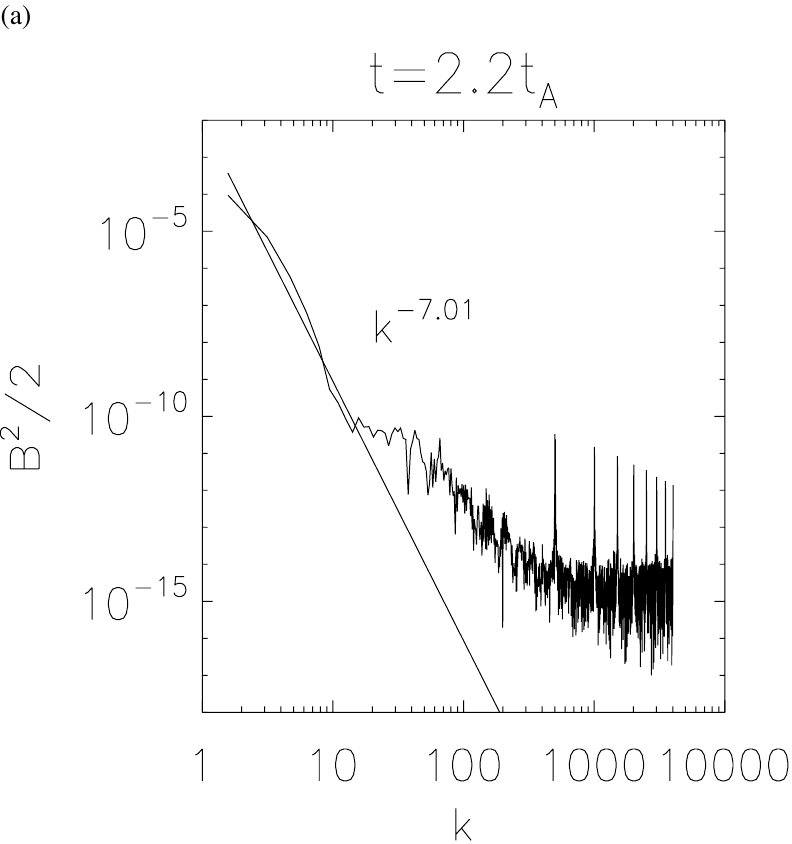}
\includegraphics[width=0.45\textwidth, clip=]  {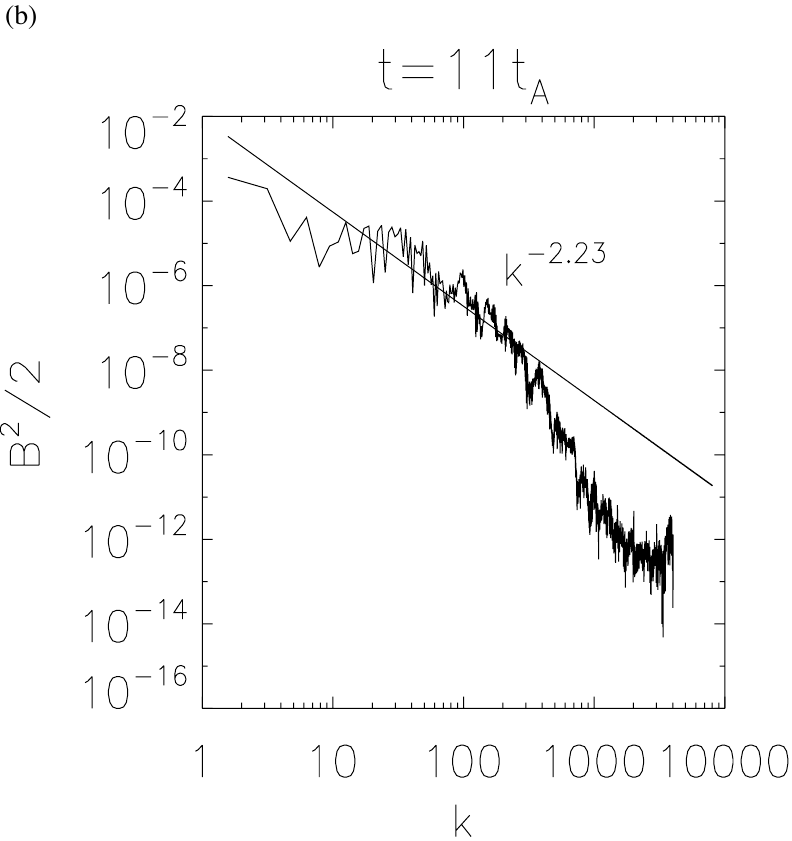}
\includegraphics[width=0.45\textwidth, clip=]  {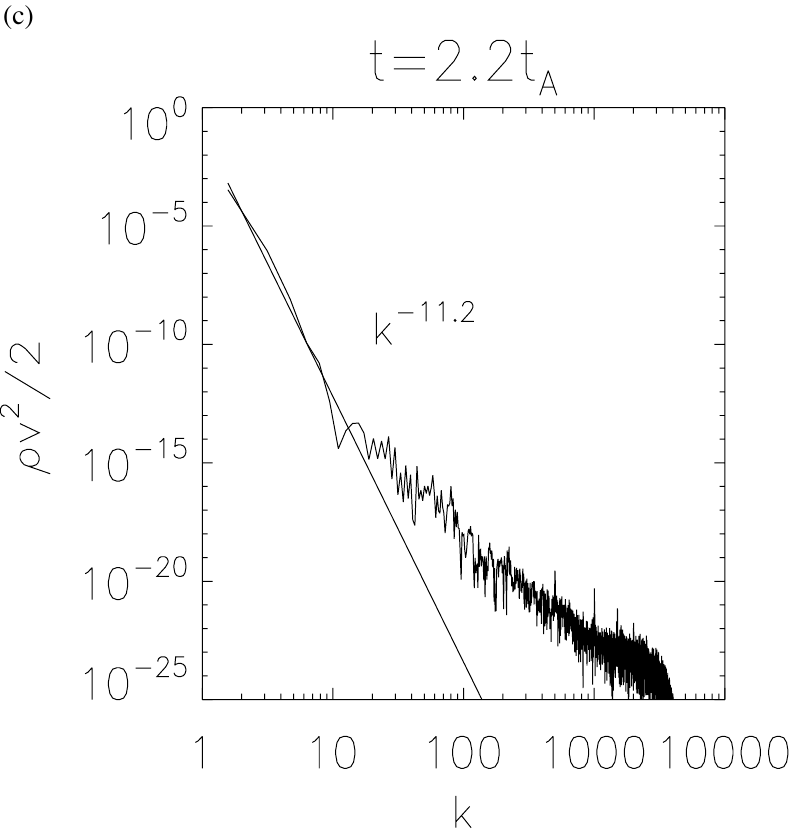}
\includegraphics[width=0.45\textwidth, clip=]  {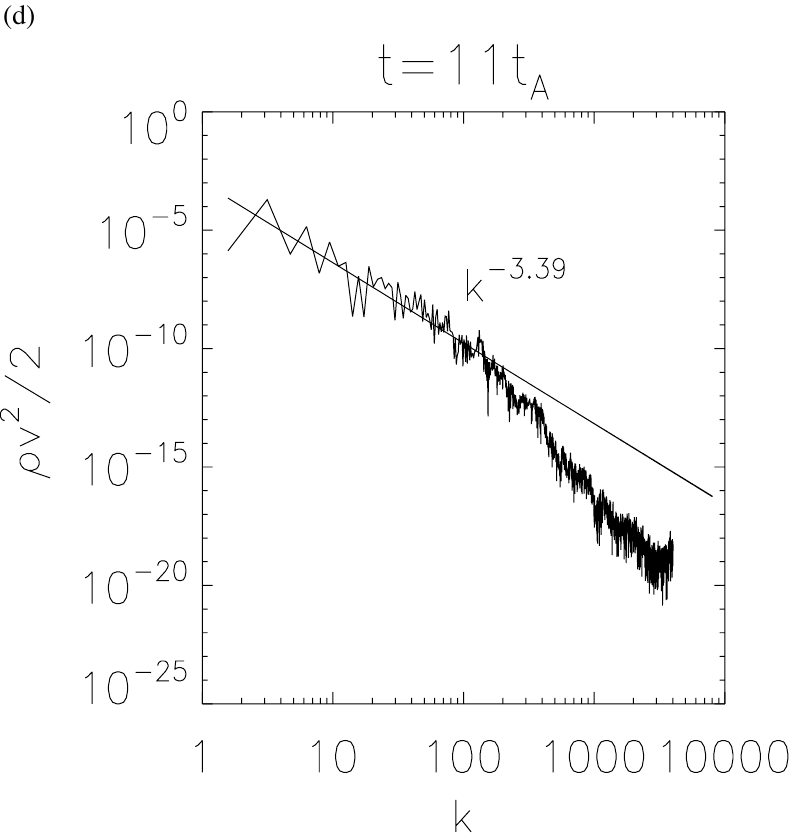}
\caption{\label{fig:5} The magnetic and kinetic energy spectrum at $t=2.2 t_A$ and $t=11t_A$ for $\beta_0=0.2$ , $S=2\times 10^5$ in model I.}
\end{figure*}

\section{Discussion and Conclusion}
Based on the 2.5 dimensional MHD numerical experiment, we have studied the effects of the initial distribution of mass density, temperature, and plasma $\beta_0$ at the inflow boundary on the plasmoid instability during magnetic reconnection. The standard Harris sheet profile was used to establish  an initial configuration in equilibrium, and a small perturbation makes the Harris sheet unstable and evolve to a thinner Sweet-Parker current sheet with shearing flows. Different from our previous paper, uniform and non-uniform guide fields are included in three different models with different initial mass density and temperature profiles. The main results and conclusions are: (1) No matter whether the Lundquist number is high or low, the non-uniform distribution of the mass density along the $x$-direction that is vertical to the current sheet strongly affects the plasmoid instability process. The high plasma mass density gradient from the center to the inflow boundary can result in a low plasma $\beta_0$ at the inflow upstream region. The higher the plasma density at the center is , the lower the plasma $\beta_0$ is.  For this kind of initial density profile, we find that : (a) For the same Lundquist number, the magnetic reconnection rate is lower for the lower $\beta_0$ case. As we have pointed out in our previous paper\cite{Ni2012}, when the low $\beta_0$ system is disturbed by the initial perturbations, it might be more difficult to push the dense plasma in the current sheet to the downstream region, leading to a slower reconnection rate.  (b) As the Lundquist number is high enough, the secondary instability appears earlier, the secondary current sheets are thinner and the current density at the main X-point is higher for a higher $\beta_0$ case. (2) When the initial density profile is uniform in the whole magnetic reconnection domain,  the different plasma $\beta_0$ at the inflow boundary corresponds to the different temperature profile. Opposite to the non-unform initial density profile model,  for the same low Lundquist number, the magnetic reconnection rate in the lower $\beta_0$ case can increase to a higher valuer, which agrees well with the analysis results of Hesse $et$ $al$.\cite{Hesse2011} and Birn $et$ $al$.\cite{Birn2011}. Increasing of the Lundquist number weakens the effects of the non-uniform initial temperature distribution on the plasmoid instability. As secondary instability appears, the effects of  $\beta_0$ induced by the different non-uniform temperature distributions can be ignored. (3) Without considering the particle acceleration, the impact of the property of the guide field on the plasmoid instability process is not apparent on the MHD scale, which is very different from the results of the kinetic simulations\cite{Drake2006}. In 2.5D resistive MHD, the guide field acts mostly like an additional pressure term. For cases with nonuniform initial mass density (such as our previous paper\cite{Ni2012} and this one), plasma pressure effects do change the onset and dynamics of the plasmoid instability, but the effects do not lead to any significant qualitative differences. There are quantitative differences including the onset criterion. However, for the case of uniform initial mass density, plasma pressure effects do not even lead to many significant quantitative differences for high Lundquist number cases ($S \ge 2000$). Kinetic simulations include the Hall effect, which can lead to symmetry breaking and current sheet tilting and similar effects for islands. Consequently, adding in a guide field to kinetic simulations leads to qualitatively different results. Additionally, particle-in-cell simulations are generally somewhat noisy, which may contribute to more efficient island formation.

In the 2.5 resistive MHD, the guide field acts mostly like an additional pressure term. For the cases with nonuniform initial mass density, our previous paper\cite{Ni2012} and this one,  do show that plasma pressure effects do change the onset and dynamics of the plasmoid instabiity, but the effects don't lead to many qualitative differences. The onset criterion is changed and there are quantitative differences. However, for the case with uniform mass density, plasma pressure effects can not even lead to many quantitative effects for the high Lundquist number case ($S \ge 2000$). Probably the most important new effect in kinetic simulations is the Hall effect. The Hall effect can lead to symmetry breaking for current sheets so that they end up tilting, and can also have similar effects on islands. The  important thing is that adding in a guide field to kinetic simulations leads to qualitatively different results. Additionally, PIC simulations are generally somewhat noisy, which may contribute to more efficient island formation. 

Though the guide fields are included, the simulations are basically carried out in the two dimensional space.  As we know, the topological structure of the magnetic reconnection process  in the real three dimensional space is very different from that in 2.5 dimensional. The present work should be re-studied in three dimensional space in the future.

\begin{acknowledgments}
This research is supported by NSFC (Grant No. 11147131), NSFC(Grant No. 11273055), NSFY(Grant No. 11203069),  the 973 program (Grant No. 2011CBB11403), CAS (Grant No. KJCX2-EW-T07 ), CAS(Grant No. 2010Y2JB16) and the Yunnan Province (Grant No.  2011FB113). N.A.M. and J. L. also acknowledge support from NSF SHINE grant AGS-1156076. This work used the NIRVANA code v3.4 developed by Udo Ziegler at the Leibniz-Institut f\"ur Astrophysik Potsdam. Calculations in this work were completed with the help of the HPC Center, Kunming Institute of Botany, CAS.
\end{acknowledgments}

\end{document}